\documentclass[aps,prd,twocolumn,groupedaddress,amssymb,eqsecnum,showpacs,epsfig]{revtex4}
\usepackage{graphicx}
\usepackage{bm}
\usepackage{dcolumn}
\usepackage{amsmath}
\def \lleq {\lower0.9ex\hbox{ $\buildrel < \over \sim$} ~}
\def \ggeq {\lower0.9ex\hbox{ $\buildrel > \over \sim$} ~}

\def \beq  {\begin{equation}}
\def \eeq  {\end{equation}}
\def \ber  {\begin{eqnarray}}
\def \eer  {\end{eqnarray}}

\def\apj{{Astroph.\@ J.\ }}

\def\aj{{Astron.\@ J.\ }}

\begin{document}
\newcommand{\newc}{\newcommand}

\newc{\be}{\begin{equation}}
\newc{\ee}{\end{equation}}
\newc{\ba}{\begin{eqnarray}}
\newc{\ea}{\end{eqnarray}}
\newc{\bea}{\begin{eqnarray*}}
\newc{\eea}{\end{eqnarray*}}
\newc{\D}{\partial}
\newc{\ie}{{\it i.e.} }
\newc{\eg}{{\it e.g.} }
\newc{\etc}{{\it etc.} }
\newc{\etal}{{\it et al.}}
\newcommand{\nn}{\nonumber}
\newc{\ra}{\rightarrow}
\newc{\lra}{\leftrightarrow}
\newc{\lsim}{\buildrel{<}\over{\sim}}
\newc{\gsim}{\buildrel{>}\over{\sim}}
\title{Crossing the Phantom Divide Barrier with Scalar Tensor Theories}
\author{L. Perivolaropoulos}
\email{http://leandros.physics.uoi.gr} \affiliation{Department of
Physics, University of Ioannina, Greece\\e-mail:\
leandros@cc.uoi.gr}
\date{\today}

\begin{abstract}
There is accumulating observational evidence (based on SnIa data)
that the dark energy equation of state parameter $w$ may be
evolving with time and crossing the phantom divide barrier $w=-1$
at recent times. The confirmation of these indications by future
data would indicate that minimally coupled quintessence can not
reproduce the observed expansion rate $H(z)$ for any scalar field
potential. Here we explicitly demonstrate that scalar tensor
theories of gravity (extended quintessence) can predict crossing
of the phantom divide barrier. We reconstruct phenomenologically
viable scalar-tensor potentials $F(\Phi)$ and $U(\Phi)$ that can
reproduce a polynomial best fit expansion rate $H(z)$ and the
corresponding dark energy equation of state parameter $w(z)$
crossing the $w=-1$ line. The form of the reconstructed scalar
tensor potentials is severely constrained but is not uniquely
determined. This is due to observational uncertainties in the form
of $H(z)$ and also because a single observed function $H(z)$ does
not suffice for the reconstruction of the two potential functions
$F(\Phi)$ and $U(\Phi)$.
\end{abstract}
\maketitle

\section{Introduction}

Detailed observations of the relation between luminocity distance
and redshift for extragalactic Type Ia Supernovae (SnIa)
indicates\cite{snobs,Riess:2004nr} that the universe has entered a
phase of accelerating expansion (the scale factor obeys ${\ddot
a}>0$). In addition, a diverse set of other cosmological data
which includes large scale redshift surveys \cite{lss} and
measurements of the cosmic microwave background (CMB) temperature
fluctuations spectrum \cite{cmb} has shown that the spatial
geometry of the universe is flat but there is not enough matter
density to justify this flatness. Thus there is an additional
cosmological component required to justify the observed flatness.
This component should have repulsive gravitational properties to
justify the observed present accelerated expansion. Such
properties can either be due to a modified theory of
gravity\cite{modgrav} or, in the context of standard general
relativity, to the existence of a smooth energy component with
negative pressure termed `dark energy'\cite{dark
energy,Sahni:2004ai}. This component is usually described by an
equation of state parameter $w\equiv{p\over \rho}$ (the ratio of
the homogeneous dark energy pressure $p$ over the energy density
$\rho$). For cosmic acceleration, a value of $w<-{1\over 3}$ is
required as indicated by the Friedmann equation \be {{\ddot
a}\over a}=-{{4\pi G}\over 3}(\rho +3p) \label{fried}\ee

The cosmological constant ($w=-1$) is the simplest viable
candidate for dark energy. It predicts an expansion history of the
universe which is described by a Hubble parameter $H(z)$ as a
function of the redshift $z$ given by \be \label{lcdm}
H^2(z;\Omega_{0m}) = \left({{{\dot a}}\over a}\right)^2 = H_0^2
[\Omega_{0m} (1+z)^3 + (1- \Omega_{0m})] \ee where flatness has
been imposed and $\Omega_{0m}\equiv {\rho_0}/{\rho_c}$ is the
single free parameter of this simplest data consistent
parametrization (LCDM). Such a model has two important
disadvantages:
\begin{itemize}
\item It requires extreme fine tuning of the value of the
cosmological constant $\Lambda$ for the accelerating expansion to
start at around the present cosmological time (coincidence
problem). \item It does not provide the best possible fit to
recent SnIa data.
\end{itemize}
In particular, recent
analyses\cite{phant-obs2,alam1,nesper,Lazkoz:2005sp} of the latest
and most reliable SnIa dataset (the Gold
dataset\cite{Riess:2004nr}) have indicated that significantly
better fits compared to LCDM are obtained by allowing for a
redshift dependent equation of state parameter. Extensive analyses
of such parametrizations of $H(z)$ have shown that the
parametrizations that allow crossing of the $w=-1$ line (known
also as Phantom Divide Line - PDL) provide significantly better
fits to the data. It is therefore important to construct physical
models that allow for a redshift-dependent $w$ that crosses the
PDL. It has been shown however that this is a highly non-trivial
task\cite{Vikman:2004dc,Perivolaropoulos:2004yr,McInnes:2005vp}.

The simplest approach towards constructing a physical model for
dark energy is to associate it with a homogeneous minimally
coupled scalar field $\phi$ with negative pressure whose dynamics
is determined by a potential properly chosen so that the energy
density of $\phi$ comes to dominate the universe at the present
time. Such models are described by Lagrangians of the form \be
{\cal L}= \pm \frac{1}{2} {\dot \phi}^2 - V(\phi) \label{lag1} \ee
where the upper (lower) sign corresponds to a
quintessence\cite{quintess} (phantom\cite{phantom}) field. It
should be noted however that phantom fields\cite{phantom}, with
negative kinetic term ($w<-1$), violate the strong energy
condition, the null energy condition ($\rho + p\geq 0$) and maybe
physically unstable. They also lead to a future `Big Rip'
singularity where all bound systems get dissociated due to the
increasing repulsive gravitational force of phantom
energy\cite{Caldwell:2003vq}. The phantom  instability however can
be cured in extended gravity theories\cite{Carvalho:2004ty}. The
equation of state parameter corresponding to (\ref{lag1}) is \be
w=\frac{p}{\rho} = \frac{\pm \frac{1}{2} {\dot \phi}^2 -
V(\phi)}{\pm \frac{1}{2} {\dot \phi}^2 + V(\phi)} \label{eqst1}
\ee For quintessence (phantom) models with $V(\phi)
> 0$ ($V(\phi) < 0$) the parameter $w$ remains in the range $-1 <
w < 1 $. For an arbitrary sign of $V(\phi)$ the above restriction
does not apply but it is still impossible for $w$ to cross the
 PDL  $w=-1$ in a continous manner.
The reason is that for $w=-1$ a zero kinetic term $\pm {\dot
\phi}^2 $ is required and the continous transition from $w<-1$ to
$w>-1$ (or vice versa) would require a change of sign of the
kinetic term. The sign of this term however is fixed in both
quintessence and phantom models. This difficulty in crossing the
PDL $w=-1$ could play an important role in identifying the correct
model for dark energy in view of the fact that data favor $w\simeq
-1$ and furthermore parametrizations of $w(z)$ where the PDL is
crossed appear to be favored over the cosmological constant
$w=-1$. Even for generalized k-essence
Lagrangians\cite{Armendariz-Picon:2000ah,Melchiorri:2002ux} of a
minimally coupled scalar field eg \be {\cal L}=\frac{1}{2}f(\phi)
{\dot \phi}^2 - V(\phi) \label{crpdl2} \ee it has been shown
\cite{Vikman:2004dc} to be impossible to obtain crossing of the
PDL. Multiple field Lagrangians (combinations of phantom with
quintessence
fields\cite{Guo:2004fq,Caldwell:2005ai,Hu:2004kh,Stefancic:2005cs})
have been shown to in principle achieve PDL crossing but such
models are complicated and without clear physical motivation (but
see \cite{Nojiri:2005vv} for an interesting physically motivated
model).

The obvious class of theories that could lead to a solution of the
above described problem is the non-minimally coupled scalar
fields. Such theories are realized in a universe where gravity is
described by a scalar-tensor theory and their study is well
motivated for two reasons:
\begin{enumerate}
\item A scalar-tensor theory of gravity is predicted by all
fundamental quantum theories that involve extra dimensions. Such
are all known theories that attempt to unify gravity with the
other interactions (eg supergravity (SUGRA), M-theory etc). \item
As shown in the following sections, quintessence scalar fields
emerging from scalar tensor theories (extended quintessence) can
predict an expansion rate $H(z)$ that violates the inequality \be
\frac{d(H(z)^2/H_0^2)}{dz}\geq 3 \Omega_{0m} (1+z)^2 \label{ineq1}
\ee It is easy to show that the inequality (\ref{ineq1}) is
equivalent to  \be \label{wz3} w(z)={{p_{DE}(z)}\over
{\rho_{DE}(z)}}={{{2\over 3} (1+z) {{d \ln H}\over {dz}}-1} \over
{1-({{H_0}\over H})^2 \Omega_{0m} (1+z)^3}}\geq -1 \ee (see eg
Ref. \cite{nesper} for a derivation of $w(z)$ in the form of
equation (\ref{wz3})). Thus, violation of the inequality
(\ref{ineq1}) is equivalent to crossing the PDL $w=-1$ and is
favored by the Gold SnIa dataset
\cite{Lazkoz:2005sp,Hannestad:2004cb,Alam:2004ip,Feng:2004ad,Alam:2004jy,Bento:2004ym}.
The inequality (\ref{ineq1}) can not be violated in minimally
coupled quintessence theories (see eq. (\ref{fzmc})).
\end{enumerate}
The usual
approach\cite{Perivolaropoulos:2004yr,Colistete:2004fv,Alcaniz:2005wk,Bento:2004ym,Wang:2004nm,Kallosh:2003bq}
in comparing quintessence models with observations is to start
with a well defined theoretical model Lagrangian, identify the
predicted form of $H(z)$ and compare with observational data to
identify the consistency and the quality of fit of the model.

This approach is not particularly efficient in view of the
infinite number of possible model Lagrangians that may be
considered. An alternative, more efficient approach is to start
from the best fit parametrization $H(z)$ obtained directly from
data and use this $H(z)$ to reconstruct the corresponding
theoretical model Lagrangian. This later approach was pioneered in
Refs \cite{Starobinsky:1998fr,Nakamura:1998mt,Huterer:1998qv} and
has been further developed for the cases of both minimally coupled
quintessence \cite{Saini:1999ba,Chiba:2000im,Rubano:2001xi} and
scalar tensor
theories\cite{Boisseau:2000pr,Esposito-Farese:2000ij} (extended
quintessence\cite{Perrotta:1999am,Uzan:1999ch,Amendola:1999qq,Chiba:1999wt}).
Extensions of this approach have recently also been applied to
$f(R)$ generalized gravity theories\cite{Capozziello:2005ku}.
However, despite the high quality of the data of the Gold dataset
that allows a fairly reliable determination of $H(z)$ (especially
at redshifts up to $z\simeq 1$) no attempt has been made to
reconstruct quintessence or extended quintessence Lagrangians from
the $H(z)$ that best fits the Gold dataset. This task is
undertaken in the present study.

We consider a simple polynomial parametrization of $H(z)$ and fit
it to the Gold dataset following Refs.
\cite{Alam:2004jy,Lazkoz:2005sp,Alam:2004ip}. The best fit form of
$H(z)$ is found to violate the inequality (\ref{ineq1}) ie to
cross the PDL. Thus it is not consistent with minimally coupled
quintessence. We thus use it to reconstruct a class of extended
quintessence scalar-tensor Lagrangian which involves two scalar
functions $F(\Phi)$ and $U(\Phi)$.

The structure of the paper is the following: In section II we
present in some detail the reconstruction method described above
for the case of extended quintessence and derive the main
equations that relate the scalar-tensor potentials $F(\Phi)$ and
$U(\Phi)$ with the observed Hubble parameter $H(z)$ and other
experimental constraints. In section III we apply this technique
to reconstruct a class of $F(\Phi)$ and $U(\Phi)$ from the actual
best fit polynomial $H(z)$ obtained from the Gold dataset. The
reconstructed potentials are consistent with other observational
constraints (solar system tests and Cavendish type experiments).
We show that in contrast to minimally coupled quintessence, this
reconstruction is possible for extended quintessence. However,
even though the reconstructed potentials $F(\Phi)$ and $U(\Phi)$
are severely constrained by the form of $H(z)$ and other
consistency requirements, they are not uniquely determined.
Finally, in section IV we review the prospects for a more
constrained reconstruction of scalar-tensor theories which may
significantly contribute towards the identification of the
fundamental theory realized in Nature.

\section{Technique for reconstructing the Scalar-Tensor Lagrangian}
The scalar-tensor Lagrangian is a generalization of the general
relativistic Lagrangian of the form \be {\cal
L}=\frac{F(\Phi)}{2}~R - \frac{Z(\Phi)}{2}~g^{\mu\nu}
\partial_{\mu}\Phi
\partial_{\nu}\Phi
- U(\Phi)  + {\cal L}_m[\psi_m; g_{\mu\nu}]\ . \label{ljf} \ee
where we have set $8\pi G=1$ ($F_0=1$) and ${\cal L}_m$ represents
the matter fields and does not depend on $\Phi$ so that the weak
equivalence principle is satisfied. One of the degrees of freedom
$F(\Phi)$ and $Z(\Phi)$ can be absorbed by a redefinition of
$\Phi$. The following parametrizations are commonly used
\begin{itemize}
\item $F(\Phi)\rightarrow \Phi$, $Z(\Phi)\rightarrow
\frac{\omega(\Phi)}{\Phi}$ is the Brans-Dicke (BD) parametrization
where $\omega(\Phi)=\frac{F(\Phi)}{(dF/d\Phi)^2}$ is the BD
parameter. \item $Z(\Phi)\rightarrow \pm 1$. The case
$Z\rightarrow -1$ corresponds\cite{Boisseau:2000pr} to a BD
parameter $-3/2<\omega_0 <0$ which contradicts solar system
tests\cite{will-bounds} for any $U(\Phi)$ allowing cosmological
evolution of $\Phi$. \item $g_{\mu \nu}\rightarrow g^*_{\mu\nu}
\equiv F(\Phi)~g_{\mu\nu}$ and $\Phi\rightarrow \varphi
:\left({d\varphi\over d\Phi}\right)^2 \equiv {3\over 4}\left({d\ln
F(\Phi)\over d\Phi}\right)^2 + {Z(\Phi)\over 2F(\Phi)}$. This
transformation corresponds to the Einstein frame as opposed to the
original Jordan frame of equation (\ref{ljf}). In the Einstein
frame the kinetic terms of the graviton and the scalar field are
diagonalized and the mathematical analysis of the theory is
simplified at the expence of introducing an explicit coupling of
the scalar field with matter. An advantage of the original Jordan
frame is that the various physical quantities are those measured
in experiments.
\end{itemize}
In what follows we choose the Jordan frame and the parametrization
$Z\rightarrow 1$. Under this assumption, $F(\Phi)$ should satisfy
the following constraints
\begin{itemize}
\item $F(\Phi)>0$ so that gravitons carry positive
energy\cite{Boisseau:2000pr}. \item $\omega_0^{-1}=(dF/d\Phi)_0^2
< 4\times 10^{-4}$ from solar system
measurements\cite{will-bounds} (the subscript $_0$ denotes the
present time).
\end{itemize}
Assuming a homogeneous $\Phi$ and varying the action corresponding
to (\ref{ljf}) in background of a flat FRW metric \be ds^2 = -dt^2
+ a^2(t)(dr^2 + r^2 \left(d\theta^2 + \sin^2\theta~d\phi^2\right))
\label{dl2}\ee we find the coupled system of
equations\cite{Esposito-Farese:2000ij}
\begin{eqnarray}
3F\cdot H^2 &=&  \rho +{1\over 2} \dot\Phi^2 - 3 H \cdot \dot F +
U
\label{fe1}\\
-2F\cdot\dot H  &=& (\rho+p) + \dot \Phi^2 +\ddot F - H\cdot \dot
F \label{fe2} \label{dmt}
\end{eqnarray}
where we have assumed the presence of a perfect fluid $(\rho,p)$.
Eliminating $\dot \Phi^2$ from (\ref{dmt}), setting \be q(z)\equiv
H(z)^2/H_0^2\label{qz} \ee and rescaling $U\rightarrow U\cdot
H_0^2$ while expressing in terms of redshift $z$ we obtain
\begin{widetext}
\ba F'' &+& \left[\frac{q'}{2q}-\frac{4}{1+z}\right]~F' +
\left[\frac{6}{(1+z)^2} - \frac{2}{(1+z)}\frac{q'}{2q}\right]~F =
\frac{2  U}{(1+z)^2 q^2} + 3 \frac{1+z}{q^2}
\Omega_{0m}\  \label{fe1a} \\
\Phi'^2 &=& -{6 F'\over 1+z} + {6 F\over (1+z)^2} -{{2 U}\over
(1+z)^2 q^2}  - 6 \frac{1+z}{q^2}  \Omega_{0m} \label{fe2a} \ea
\end{widetext} where the prime $'$ denotes differentiation with
respect to redshift $(\frac{d}{dz})$ and we have assigned
properties of matter ($p=0,\;
\Omega_{0m}=\frac{3\rho_{0m}}{H_0^2}$) to the perfect fluid.

Given the form of $H(z)$ from observations, equations (\ref{fe1a})
and (\ref{fe2a}) may be used with boundary conditions $F(0)=1$,
$F'(0)=0$ and $\Phi(0)=0$ to reconstruct sets of $F(z)$, $U(z)$,
$\Phi(z)$ which predict the given form of $H(z)$ and are
consistent with solar system tests. The system (\ref{fe1a}) and
(\ref{fe2a}) may take a more convenient form by setting \be
F(z)=\frac{f(z)}{(1+z)^2} \label{ff} \ee It then takes the form
\begin{widetext} \ba U(z)&=&\frac{1}{2} (1+z)^4 q f''+\frac{1}{4}
q'
(1+z)^4 f' -\frac{3}{2}(1+z)^3 \Omega_{0m} \label{fe1b} \\
\Phi'(z)^2 &=& -6f-6(1+z)f'-\frac{q'}{2q} (1+z)^2 f' -(1+z)^2 f''
-3 \Omega_{0m} \frac{1+z}{q} \label{fe2b} \ea \end{widetext}with
boundary conditions $\Phi(0)=0$, $f(0)=1$ and $f'(0)=-2$.

The case of minimally coupled $\Phi$ is regained for $F(z)=1$
($f=1/(1+z)^2$). Indeed, setting $f(z)=1/(1+z)^2$ in
(\ref{fe1b})-(\ref{fe2b}) we find \ba U(z) &=&  3q
-\frac{q'}{2} (1+z)-\frac{3}{2} (1+z)^3 \Omega_{0m} \label{uzmc}\\
\Phi'^2 (z)&=& \frac{1}{q(1+z)} (q'-3\Omega_{0m} (1+z)^2)
\label{fzmc} \ea which are in agreement with previous
studies\cite{Saini:1999ba,Starobinsky:1998fr} reconstructing
minimally coupled Lagrangians.

The reconstruction of $U(\Phi)$ may proceed in the minimally
coupled case by finding $\Phi(z)$ from (\ref{fzmc}) (setting eg
$\Phi(0)=0$), inverting for $z(\Phi)$ and substituting in
(\ref{uzmc}) to find $U(\Phi)$. The inequality (\ref{ineq1})
(preventing PDL crossing), valid for the minimally coupled case is
recovered from (\ref{fzmc}) since $\Phi'^2 >0$.

As a simple application we may consider the $H(z)$ induced by a
cosmological constant which implies \be q(z)= \Omega_\Lambda +
\Omega_{0m} (1+z)^3 \label{qcc}\ee In this case
(\ref{uzmc})-(\ref{fzmc}) reduce to \ba U(z)&=&
3\Omega_\Lambda=\rho_\Lambda \\ \Phi'(z)&=&0 \ea  as expected.
Thus, the reconstruction leads to a uniquely defined potential in
the minimally coupled case if $q'>3\Omega_{0m} (1+z)^2$.

If on the other hand there is a redshift range where
$q'<3\Omega_{0m} (1+z)^2$, an $F(z)\neq 1$ ($f\neq 1/(1+z)^2$) is
required to keep $\Phi'^2>0$ in (\ref{fe2b}). Since
$q'<3\Omega_{0m} (1+z)^2$ implies superacceleration which can not
be supported merely by the repulsive gravity of the potential $U$,
a modified strength of gravity will be required ($F(z)<1$). Such
an $F(z)$ is also constrained to obey $F(z)>0$ for all redshifts
and $F'(z=0)\simeq 0$. It is therefore possible that no such
$F(z)$ exists so that for a given $q(z)$ the three requirements
\ba \Phi'(z)^2 &\geq&0 \label{req1} \\ F(z)&>&0 \label{req2}
\\F'(z=0)&=&0 \label{req3} \ea are fulfilled. In fact, for $q(z)$ corresponding
to a cosmological constant (equation (\ref{qcc})) and $U(z)=0$ it
has been shown that no acceptable function $F(z)$
exists\cite{Esposito-Farese:2000ij} because $F(z)$ becomes
negative at relatively low $z$. If an acceptable $F(z)$ is found
to exist then it will not be uniquely defined because a perturbed
$F(z)$ will also lead to a positive $\Phi'^2$ leading to another
acceptable solution.

The best case scenario for the existence of an acceptable $F(z)$
is the case when $\Phi'(z)=0$ ie when $F(z)$ does not have to
balance the attractive gravity of $\Phi'^2$. If no acceptable
$F(z)$ exists for $\Phi'=0$ (ie if $F(z)<0$ for $z>z_c$) then
there will be no acceptable $F(z)$ for any $\Phi'^2 >0$ ($F(z)$
will get even more negative to balance the larger attractive
gravity of $\Phi'^2$). Thus, in order to see if an acceptable
scalar tensor theory exists for a given $q(z)$ we simply have to
set $\Phi'=0$ in equation (\ref{fe2b}), solve for $f(z)$ with
$f(0)=1$, $f'(0)=-2$ and see if the solution obeys $f(z)>0$. If
$f(z)<0$ for some redshift range then it will remain so for any
$\Phi'^2 >0$ and therefore no acceptable scalar tensor Lagrangian
exists predicting the given $q(z)$. If on the other hand $f(z)>0$
for all redshifts of interest, the solution is acceptable and we
may start increasing $\Phi'^2$ (trying different functional forms)
until $f(z)<0$ for some redshift range. Thus we may identify a set
of observationally acceptable scalar tensor Lagrangians that
predict the given form of $q(z)$. This procedure will be applied
in the next section to identify scalar-tensor Lagrangians that
predict the particular form of $q(z)$ that consists one of the
best fits to the Gold dataset.

\section{From the Gold Dataset to the Fundamental Theory}

A wide range of expansion history parametrizations $H(z)$ with a
flat prior\cite{phant-obs2,alam1,nesper,Lazkoz:2005sp} has been
fitted to the Gold and earlier SnIa datasets. The best fits among
those parametrizations share the following common features
\begin{itemize}
\item The inequality (\ref{ineq1}) is violated for $z\lsim 0.3$ ie
the effective equation of state is $w(z)<-1$ for $z\lsim 0.3$.
\item The inequality (\ref{ineq1}) is respected for $z\gsim 0.3$
ie the PDL is crossed at $z\gsim 0.2-0.3$.
\end{itemize}
A representative easy to handle parametrization that shares these
common features at best fit and has been one of the first
introduced in the literature is the polynomial
parametrization\cite{alam1} \be q(z)=\Omega_{0m} (1+z)^3 + a_2
(1+z)^2 + a_1 (1+z) + a_0 \label{polypar} \ee where $a_0\equiv
1-a_2-a_1-\Omega_{0m}$ due to flatness. For a prior of
$\Omega_{0m}=0.3$ the parameter values that best fit the Gold
dataset are \cite{Lazkoz:2005sp} $a_1=-4.16\pm 2.53$, $a_2=1.67\pm
1.03$. The effective equation of state parameter $w(z)$
corresponding to the parametrization (\ref{polypar}) is shown in
Fig. 1 along with the corresponding form of $H(z)$ at best fit.
The form of $H(z)$ corresponding to a cosmological constant with
the same prior is also shown for comparison.
\begin{figure}[h]
\centering
\includegraphics[bb=30 140 500 780,width=6.7cm,height=8cm,angle=-90]{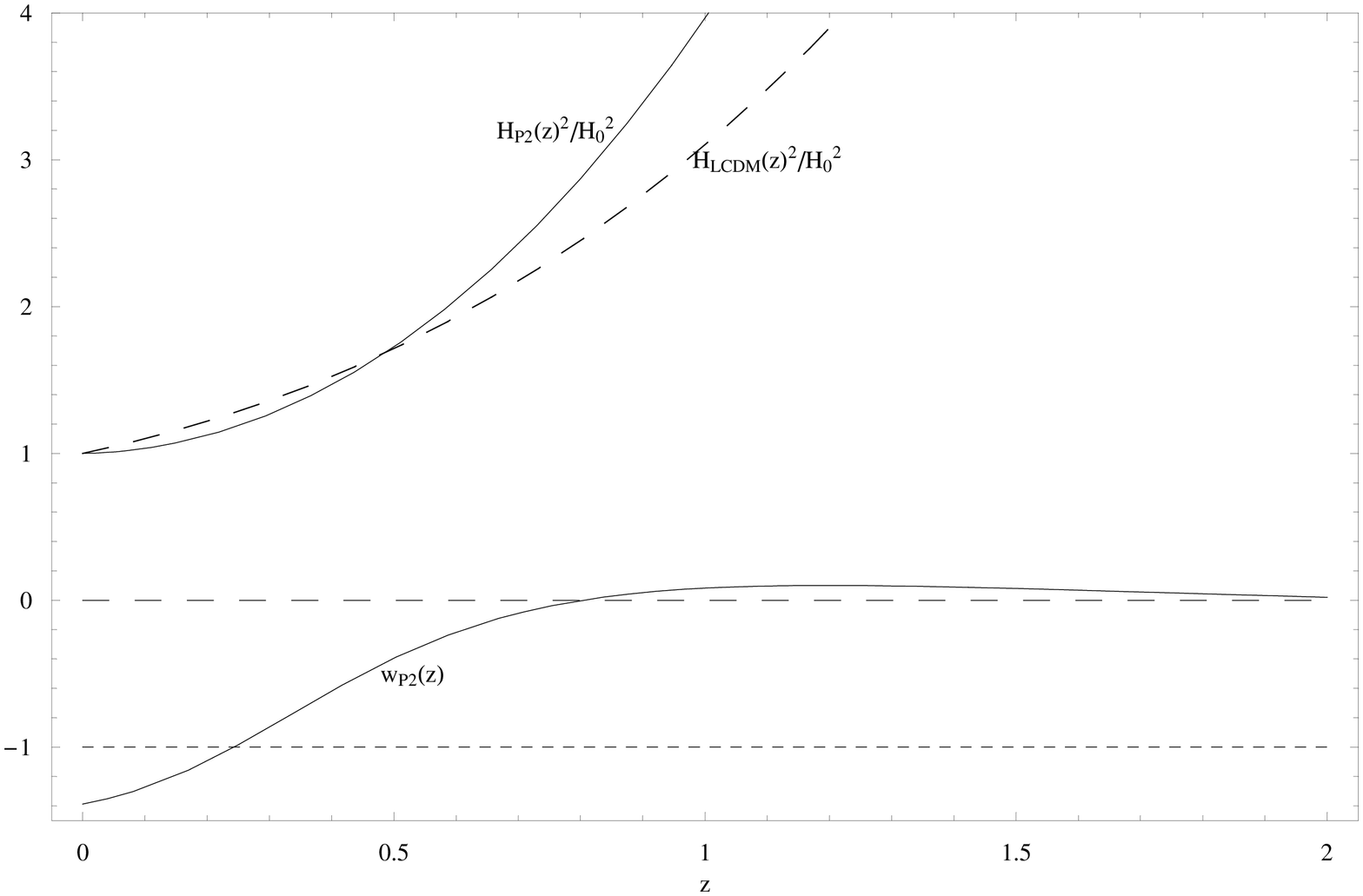}
\caption{The effective equation of state parameter $w(z)$
corresponding to the parametrization (\ref{polypar}) along with
the corresponding form of $H(z)=H_{P2}(z)$ at best fit. The form
of $H(z)=H_{LCDM}(z)$ corresponding to a cosmological constant
with the same prior is also shown for comparison (dashed line).}
\label{fig1}
\end{figure}
In the redshift range $0<z<0.25$ where $w(z)<-1$ the inequality
(\ref{ineq1}) is temporarily violated.  The PDL is clearly crossed
and therefore no minimally coupled scalar can account for the
accelerating expansion crossing $w(z)=-1$ (see equation
(\ref{fzmc})). We will thus examine if a non-minimally coupled
scalar can account for this type of expansion.

The parametrization (\ref{polypar}) will be used for the
reconstruction of the scalar tensor potentials along the lines
described in the previous section. No attempt will be made to
reconstruct the unique scalar tensor theory obtained from data.
Instead we only examine if it is in principle possible to
construct extended quintessence Lagrangians that lead to a
crossing of the PDL in the way favored by the Gold dataset. We
thus use arbitrary forms of $\Phi'(z)$  and first try the `best
case scenario' for the existence of $F(z)$ setting $\Phi'(z)=0$ in
(\ref{fe2b}) while using boundary conditions $f(0)=1$, $f'(0)=-2$.
The solution of (\ref{fe2b}) for $F(z)=(1+z)^2 f(z)$ is easily
found (see Fig. 2) to fulfil the requirement (\ref{req2})
((\ref{req3}) is imposed by the boundary conditions) for all
redshifts $z$ and is therefore observationally acceptable.
\begin{figure}[h]
\centering
\includegraphics[bb=30 100 420 700,width=6.7cm,height=8cm,angle=-90]{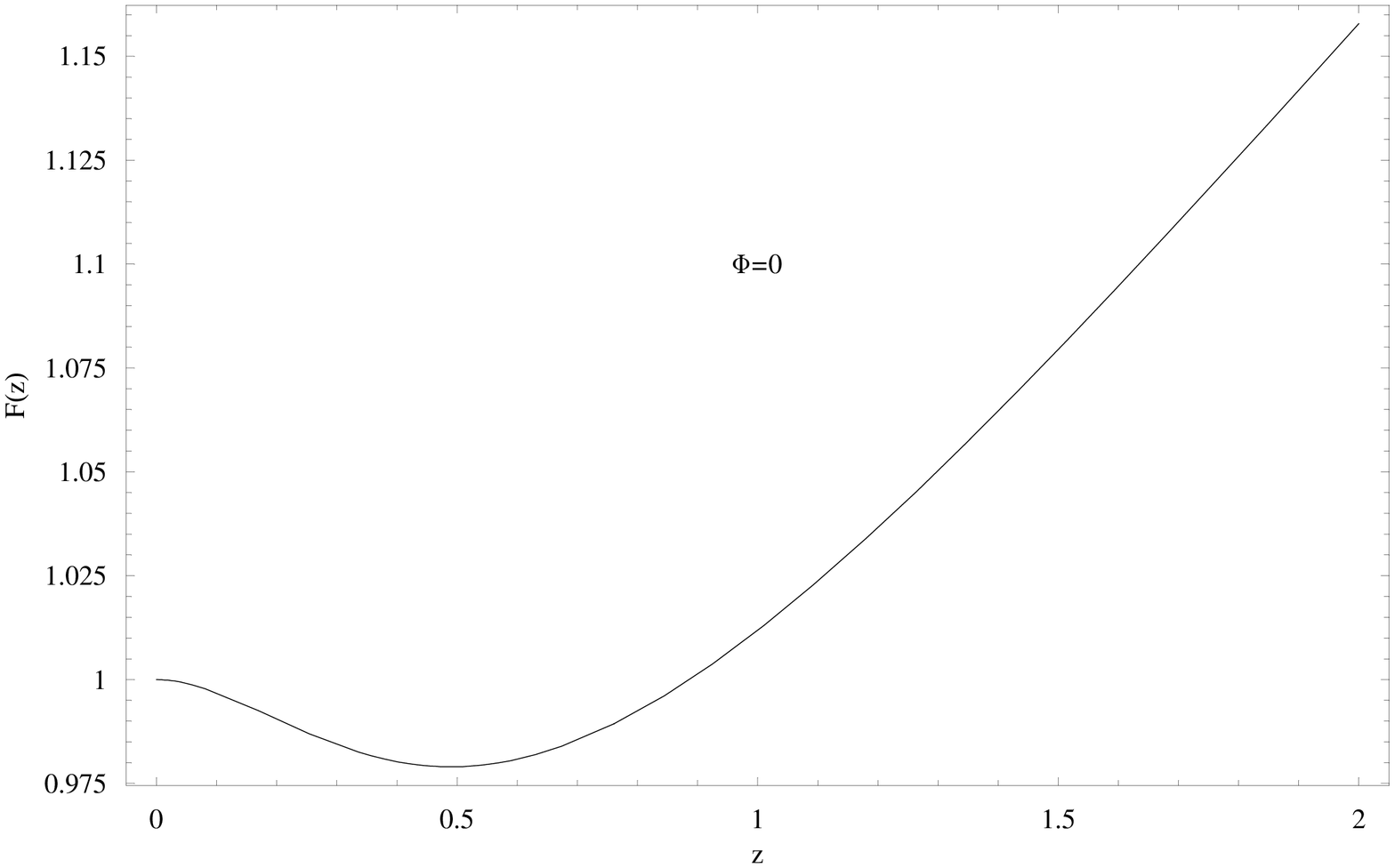}
\caption{The function $F(z)$ can fulfil the requirements
(\ref{req2})-(\ref{req3}) for a constant $\Phi(z)$ (eg $\Phi=0$).}
\label{fig2}
\end{figure}

We next ask the following question: `Can the requirements
(\ref{req1})-(\ref{req3}) continue to hold if we use a small
positive $\Phi'(z)^2$?'. By numerically solving equation
(\ref{fe2b}) for a few trial functions $\Phi'(z)^2$ and boundary
conditions $f(0)=1$, $f'(0)=-2$ it is easy to see that only small,
decreasing with $z$ functions $\Phi'(z)$ can accommodate $F(z)>0$
for all redshifts $z$. Examples of such functions include \ba
\Phi'(z)^2 &=& a\; (1+z)^{-n} \label{planz} \\
\Phi'(z)^2 &=& a\; e^{-z} \label{expanz} \ea with $a\lsim 0.35$
and $n\gsim 1$. All such functions lead to qualitatively similar
forms for $f(z)$.

In order to reconstruct the scalar potentials $U(\Phi)$, $F(\Phi)$
of the scalar tensor theory for the best fit $q(z)$ of
(\ref{polypar}) we use the following steps
\begin{enumerate}
\item Pick a function $\Phi'(z)=g(z)$ leading to $F(z)>0$ in the
redshift range of interest. Such functions are given eg by
equations (\ref{planz}) and (\ref{expanz}). Use this function and
the best fit $q(z)$ of equation (\ref{polypar}) to solve equation
(\ref{fe2b}) numerically. \item Use the resulting $f(z)$ along
with the best fit $q(z)$ in equation (\ref{fe1b}) to find $U(z)$.
\item To convert $U(z)$ to $U(\Phi)$ solve the differential
equation $\Phi'(z)=g(z)$ using the selected function $g(z)$ with
boundary condition $\Phi(0)=0$ (set eg $\Phi$ to $0$ at the
present time). Use the solution $\Phi(z)$ to invert and find
$z(\Phi)$. \item Substitute $z(\Phi)$ in the $U(z)$ found in step
2 and in $F(z)=(1+z)^2 f(z)$ found in step 1 to find
$U(z(\Phi))=U(\Phi)$ and $F(z(\Phi))=F(\Phi)$.
\end{enumerate}
Notice that the inversion $z(\Phi)$ can not be made for all values
of $\Phi$ but only for those values of $\Phi$ appearing in the
$\Phi(z)$ plot. This is a limitation because we have found that
$\Phi(z)$ tends to a constant at large $z$ (see Fig. 3) because of
large cosmic friction at early times.

The above steps have been implemented in a mathematica code
available at \cite{mathfile}. For definiteness we have assumed a
$\Phi'(z)$ of an exponentially decreasing form given in
(\ref{expanz}) with $a=0.35$. The resulting functions $\Phi(z)$,
$U(z)$ and $F(z)$ are shown in Fig. 3 in the redshift range
$0<z<2$. The potential $U(z)$ is found to be a slowly increasing
function of $z$ while $F(z)$ has a shallow minimum at $z\simeq 1$
and increases monotonically beyond that redshift. The minimum of
$F(z)$ in Fig 3 (see also Fig. 5) however, is deeper compared to
that of Fig. 2 corresponding to $\Phi'(z)=0$ because $F(z)$ has to
balance, the attractive gravity of $\Phi'(z)\neq 0$ and therefore
has to divert more from $F(z)=1$. It is easily shown that the
resulting potentials remain qualitatively similar for other
acceptable forms of $\Phi'(z)$.
\begin{figure}[h]
\centering
\includegraphics[bb=50 120 460 750,width=6.7cm,height=8cm,angle=-90]{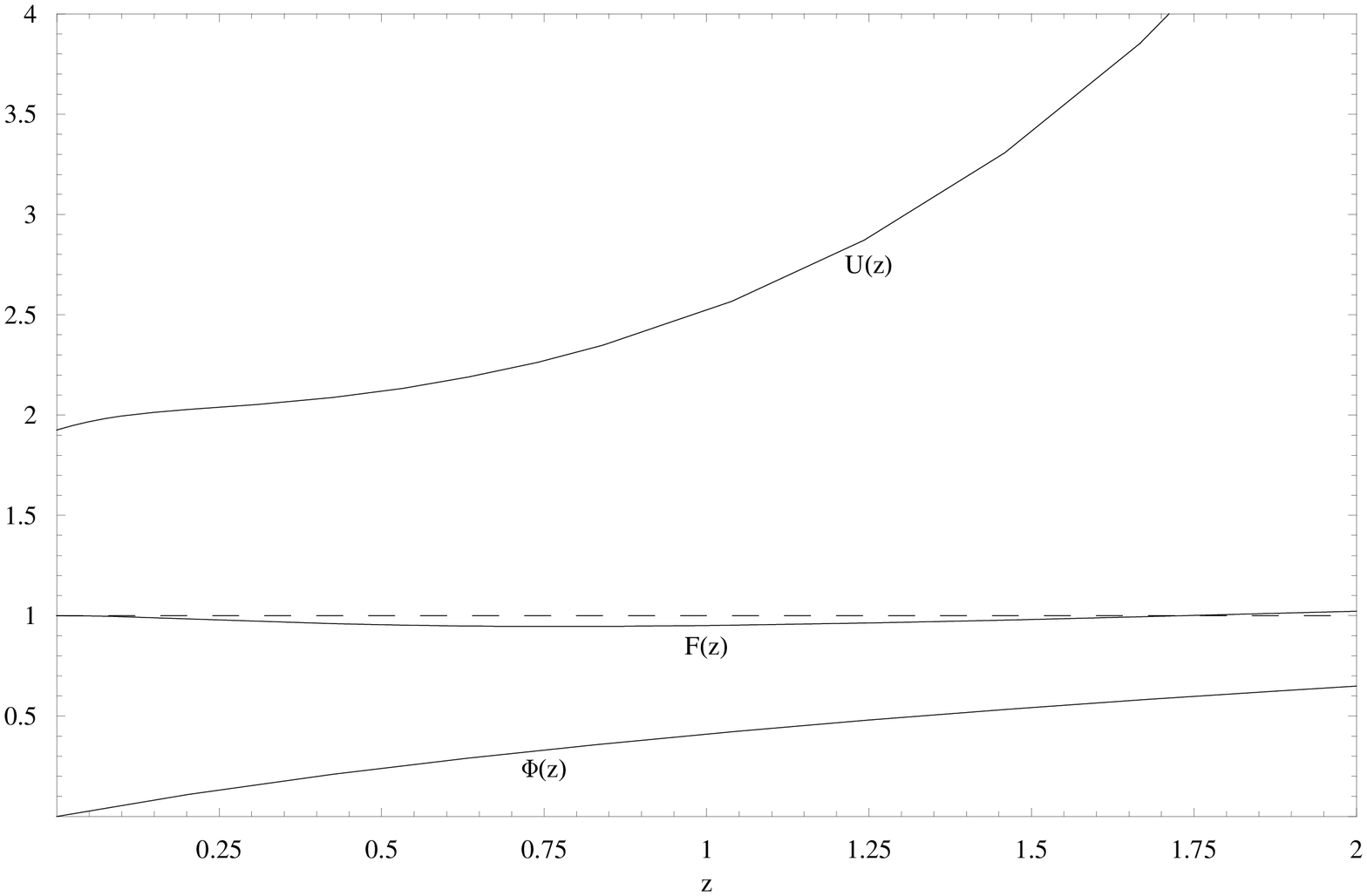}
\caption{The reconstructed functions $\Phi(z)$, $U(z)$ and $F(z)$
in the redshift range $0<z<2$.} \label{fig3}
\end{figure}
By inverting $\Phi(z)$ along the lines of step 3 the potentials
$U(\Phi)$ and $F(\Phi)$ may be found and are shown in Figs. 4 and
5 respectively. In Fig. 4 the numerically obtained potential
corresponds to the thick dotted line. The {\it existence} of such
potentials is our main result and not their particular form which
is not unique given the form of $H(z)$. Nevertheless for
illustration purposes we attempt to fit the numerically obtained
potentials with simple analytic functions. The continuous lines in
Fig. 4 correspond to attempts to fit the numerically obtained
potential with simple analytical functions. We have tried several
simple analytical functions. A very good fit was provided by an
exponential of the form \be U(\Phi)=A\; e^{\lambda \Phi^2}
\label{pu1} \ee with $\lambda \simeq 3$. The fit was worse for
other powers of $\Phi$ in the exponential (the next best
approximation $U(\Phi)\sim e^{\lambda \Phi}$ ($\lambda \simeq 5$)
is also shown in Fig. 4 and the fit is clearly not as good as with
(\ref{pu1})).
\begin{figure}[h]
\centering
\includegraphics[bb=40 120 480 780,width=6.7cm,height=8cm,angle=-90]{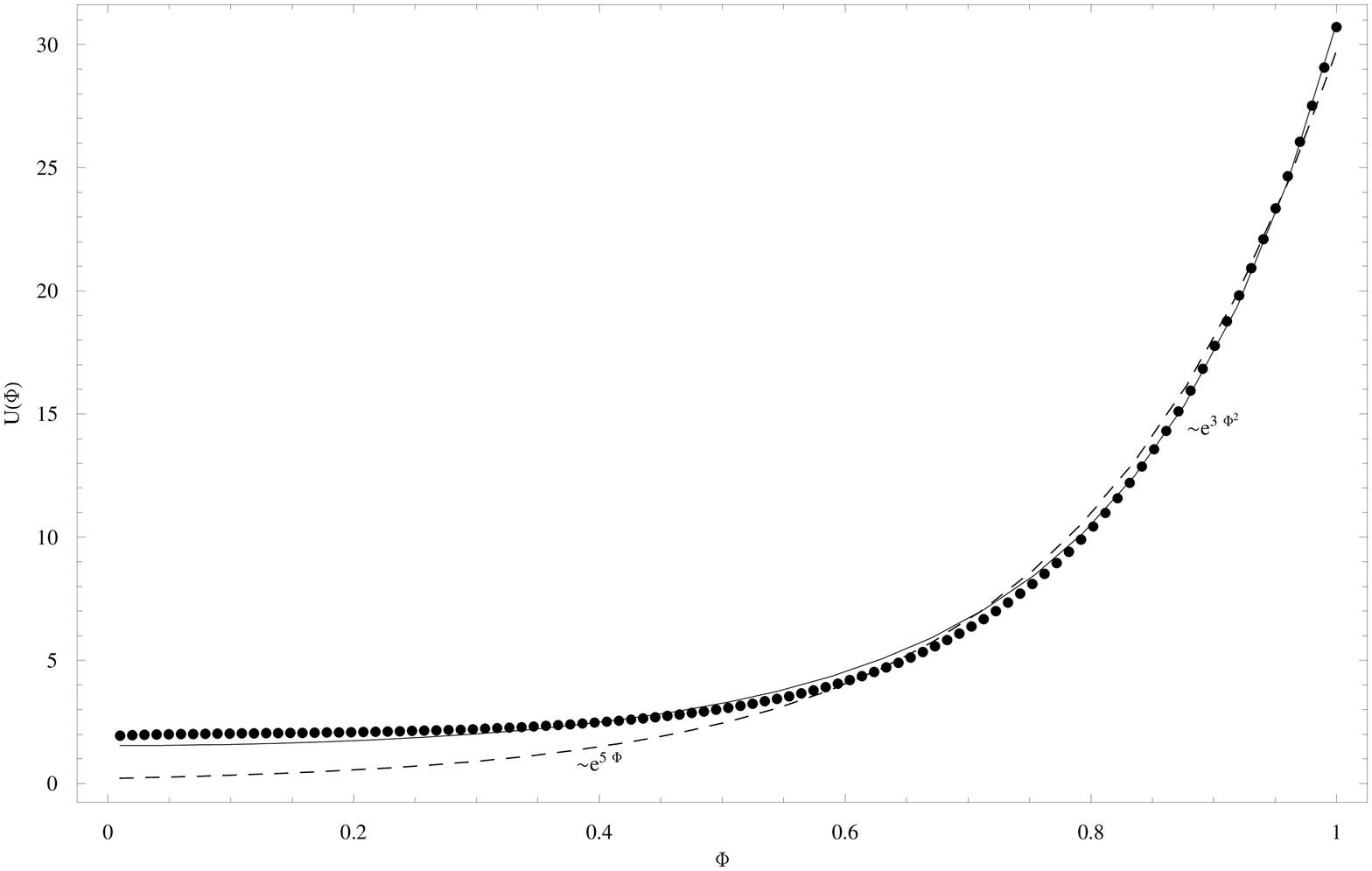}
\caption{The reconstructed potential $U(\Phi)$ is well fit by
$e^{\lambda \Phi^2}$ but not with other powers of $\Phi$ in the
exponent (eg $e^{\lambda \Phi}$).} \label{fig4}
\end{figure}

It is straightforward to test our results,  by substituting the
reconstructed functions $U(z)$ and $f(z)$ back in equation
(\ref{fe1b}) to solve for $q(z)$ numerically (with boundary
condition $q(0)=1$) and confirm that we get back the best fit form
(\ref{polypar}) with the appropriate parameter values. This test
revealed that the numerically obtained functions $U$ and $F$ of
Figs 3-5 indeed lead to a prediction of exactly the best fit form
of $q(z)$.
\begin{figure}[h]
\centering
\includegraphics[bb=40 120 480 780,width=6.7cm,height=8cm,angle=-90]{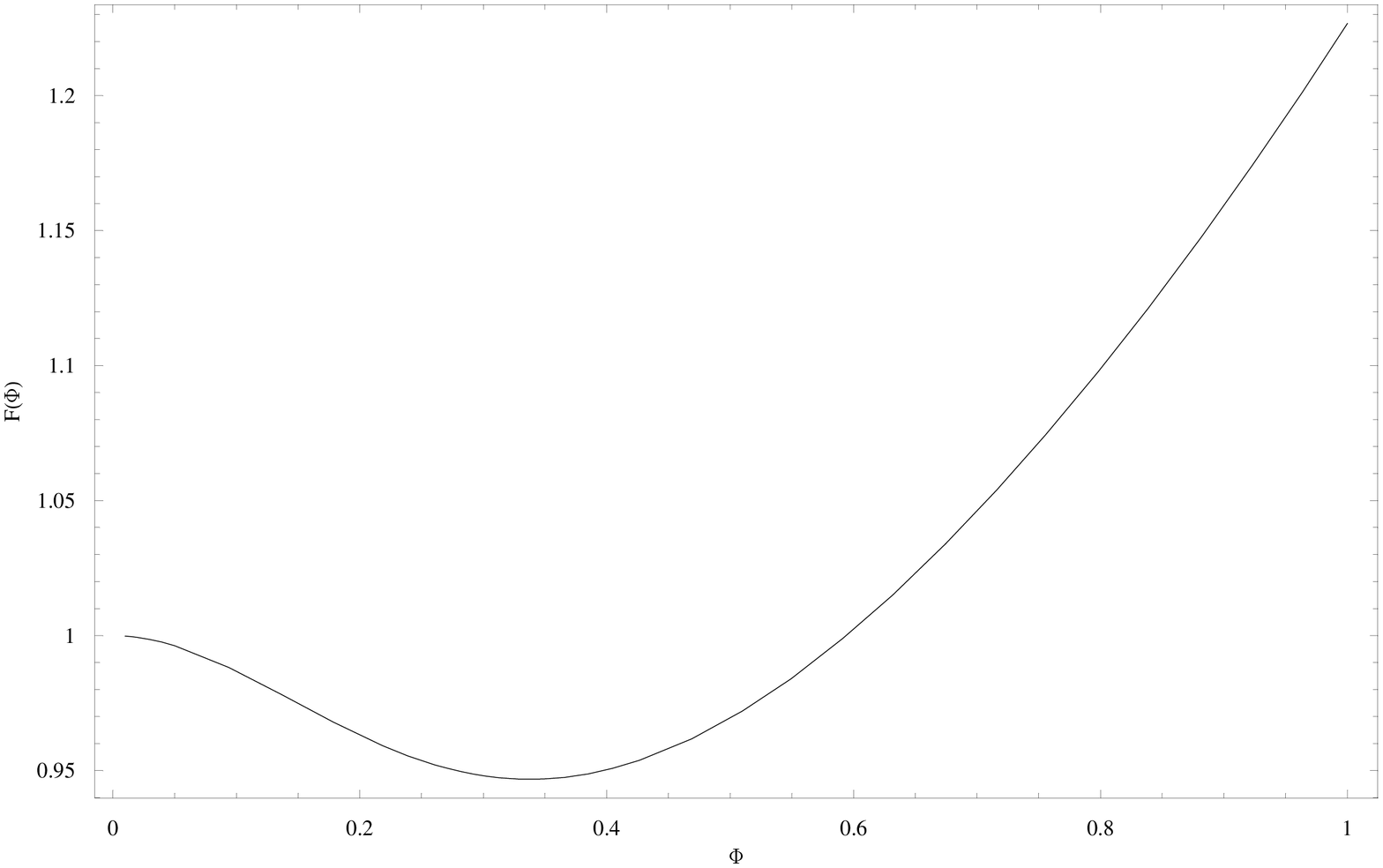}
\caption{The reconstructed function $F(\Phi)$ has a local minimum
at a small value of $\Phi$ which is deeper than that of Fig. 2.
This minimum appears to be a generic feature found using other
best fit parametrizations of $H(z)$ as well.} \label{fig5}
\end{figure}

Alternative best fit forms of $q(z)$ that violate the inequality
(\ref{ineq1}) crossing the PDL can be used to reconstruct
potentials $F(\Phi)$, $U(\Phi)$. For example Ref.
\cite{Lazkoz:2005sp} has indicated a best fit $q(z)$
parametrization that crosses the PDL repeatedly, showing evidence
for oscillations in redshift space. This parametrization was shown
to provide better fit to the Gold dataset than any other
parametrization tested in Ref. \cite{Lazkoz:2005sp}. It is
straightforward to reconstruct potentials $F$ and $U$ that
reproduce the best fit oscillating $H(z)$ parametrization and show
that the expansion history oscillations are inherited by the
reconstructed potentials\cite{mathfile}. Thus the $U\sim
e^{\lambda \Phi^2}$ form of the reconstructed potential is not a
particularly robust prediction of the Gold dataset.

\section{Conclusion}
We have demonstrated that it is possible to reconstruct
experimentally viable scalar-tensor potentials which predict
exactly the best fit parametrizations obtained from the recent
Gold SnIa dataset. These parametrizations share the common feature
of temporarily violating the inequality (\ref{ineq1}) or
equivalently correspond to a dark energy equation of state that
crosses the PDL. This feature is impossible to reproduce in the
context of single field quintessence or even phantom models.

As an application we have used a quadratic polynomial
parametrization for $H(z)$ at its best fit with respect to the
Gold dataset to reconstruct the scalar- tensor potentials
$F(\Phi)$ and $U(\Phi)$ taking into account consistency with solar
system and Cavendish type experiments. The reconstructed scalar
tensor potential $U$ was found to be well fit by an analytical
function of the form $U(\Phi)\sim e^{3\Phi^2}$ which may be
thought of as physically motivated on the basis of SUGRA
theories\cite{Caresia:2003ze,Brax:2005uf}.

The derived potentials however are not uniquely determined for two
reasons
\begin{itemize}
\item We used one observationally determined input function
($H(z)$) in the context of two coupled equations to construct {\it
three} output functions ($\Phi(z)$, $F(z)$, $U(z)$). The
additional solar system test experimental constraints were not
enough to lead to a unique determination of the three output
functions. \item The best fit form of $H(z)$ depends on the type
of parametrization considered. The polynomial parametrization
considered for $H(z)$ gives a very good quality of fit to the Gold
dataset but is not unique. For example an oscillating
parametrization gives a somewhat better fit, yet the reconstructed
potentials are quantitatively different from those obtained from
the polynomial parametrization at best fit.
\end{itemize}
Thus our main result is that in contrast to minimally coupled
quintessence, scalar tensor theories can reproduce the main
features of the best fit Hubble expansion history obtained from
the Gold dataset. However, the precise determination of the scalar
tensor theory potentials requires more accurate SnIa data and
additional observational
input\cite{Chen:2004nq,Corasaniti:2004sz,Jain:2004qy,Linder:2002dt,Garriga:2003nm}
which could come eg from weak lensing or large scale structure
surveys providing the structure formation evolution history
$\frac{\delta \rho}{\rho}(z)\equiv \delta_m(z)$. Once
$\delta_m(z)$ is known from observations its time evolution
equation \be {\ddot \delta_m} + 2H {\dot \delta_m} - 4\pi G_{\rm
eff}(t)\, \rho~\delta_m\approx 0~. \label{dm} \ee may be used in
addition to the system (\ref{fe1b})-(\ref{fe2b}) to fully
determine the scalar tensor Lagrangian. The unknown functions
$F(z)$,$U(z)$ and $\Phi(z)$ enter in equation (\ref{dm}) though
the effective Newton's constant $G_{\rm eff}(z)$ measured in
Cavendish type experiments ($Force=G_{\rm eff}\frac{m_1
m_2}{r^2}$) which is connected with the Newton's constant $G$
entering in the scalar tensor Lagrangian (\ref{ljf}) by
\cite{Esposito-Farese:2000ij} \be G_{\rm eff} \equiv  = {G\over F}
\left({2ZF+4(dF/d\Phi)^2\over2ZF+3(dF/d\Phi)^2}\right)\ .
\label{Geff} \ee In terms of the redshift $z$, equation (\ref{dm})
takes the form \be H^2~\delta_m'' + \left(\frac{(q^2)'}{2} -
{q^2\over 1+z}\right)\delta_m' \approx {3\over 2} (1+z) {G_{\rm
eff}(z)\over G}~\Omega_{0m}~\delta_m~. \label{dmz} \ee The use of
equation (\ref{dmz}) to supplement the system
(\ref{fe1b})-(\ref{fe2b}) and uniquely determine the functions
$F(z)$,$U(z)$ and $\Phi(z)$ subject only to the observational
uncertainties of the observed input functions $H(z)$ and
$\delta_m(z)$ is a particularly interesting future prospect.

\vspace{0.3cm} The Mathematica\cite{wolfram} file including the
numerical analysis and the production of the figures of the paper
can be downloaded along with the source files of the paper from
the astro-ph archive or sent by e-mail upon request.

{\bf Acknowledgements:} This research was funded by the program
PYTHAGORAS-1 of the Operational Program for Education and Initial
Vocational Training of the Hellenic Ministry of Education under
the  Community Support Framework and the European Social Fund.

\end{document}